\documentclass[referee]{mn2e}
\usepackage{graphicx}
\begin{document}
\title{On the relativistic magnetic reconnection}
\author[Y.E.Lyubarsky]{Y.E.Lyubarsky\\
Physics Department, Ben-Gurion University, P.O.B. 653, Beer-Sheva
84105, Israel; e-mail: lyub@bgumail.bgu.ac.il}
\date{Received/Accepted}
\maketitle
\begin{abstract}
Reconnection of the magnetic lines of force is considered in case
the magnetic energy exceeds the rest energy of the matter. It is
shown that the classical Sweet-Parker and Petschek models are
generalized straightforwardly to this case and the reconnection
rate may be estimated by substituting the Alfven velocity in the
classical formulas by the speed of light. The outflow velocity in
the Sweet-Parker configuration is mildly relativistic. In the
Petschek configuration, the outflow velocity is ultrarelativistic
whereas the angle between the slow shocks is very small. Due to
the strong compression, the plasma outflow in the Petschek
configuration may become strongly magnetized if the reconnecting
fields are not exactly antiparallel.
\end{abstract}
\begin{keywords}
magnetic fields -- MHD --- shock waves
\end{keywords}

\section{Introduction}
In highly conducting plasmas, the magnetic energy is released by
reconnection of the magnetic lines of force near the singular
lines where the magnetic  field changes sign. This process has
been intensively studied both in laboratory plasma devices and in
space (see recent monographs by Biskamp (2000) and Priest \&
Forbes (2000)). Magnetic reconnection may occur in relativistic
objects such as pulsars, magnetars, active galactic nuclei or
gamma-ray bursts. In pulsars, the relativistic magnetic
reconnection was proposed as a source of the high-energy emission
(Lyubarskii 1996; Kirk, Skj{\ae}raasen \& Gallant 2002) and as the
solution to the $\sigma$-problem (Coroniti 1990; Lyubarsky \& Kirk
2001; Kirk \& Skj{\ae}raasen 2003; Lyubarsky 2003). Similar models
were also developed for the cosmological gamma-ray bursts
(Drenkhahn 2002; Drenkhahn \& Spruit 2002). The magnetic
reconnection was evoked for explanation of the rapid variability
observed in active galactic nuclei (Di Matteo 1998). The particle
acceleration in the reconnection process was proposed to operate
in radio jets (Romanova \& Lovelace 1992; Birk, Crusius-W\"atzel
\& Lesch 2001 Jaroschek, Lesch \& Treumann 2004). The reconnection
of the superstrong magnetic field is a key element of the widely
recognized model for the soft gamma-repeaters (Thompson \& Duncan
1995; Lyutikov 2003). Therefore relativistic generalization of the
classical reconnection models is in order.

Blackman \& Field (1994) considered kinematics of relativistic
reconnection in the Sweet-Parker and Petschek configurations and
concluded that due to the Lorentz contraction, the reconnection
rate is significantly enhanced and may approach the speed of
light. Lyutikov \& Uzdensky (2002) confirmed this conclusion for
the Sweet-Parker case. In both these works, the full energy and
momentum balance was not considered, the authors imposed instead
condition of incompressibility assuming that the proper density of
the plasma remains constant. Particle acceleration in relativistic
current sheets was studied both in the test particle approximation
(Romanova \& Lovelace 1992; Birk, Crusius-W\"atzel \& Lesch 2001;
Larrabee, Lovelace \& Romanova 2003, Kirk 2004) and in
two-dimensional PIC simulations (Zenitani \& Hoshino 2001;
Jaroschek et al 2004).

Here we present generalization of the Sweet-Parker and Petschek
models to the relativistic case. In our analysis we will follow
the classical approach just substituting the nonrelativistic
expressions for the fluxes of the conserved quantities by the
relativistic ones. It will be shown that the outflow velocity
becomes relativistic if in the inflow region the magnetic energy
density exceeds the plasma rest energy density. The classical
formulas for the reconnection rate remain valid in this case if
one substitutes the Alfven velocity by the speed of light. The
relativistic Petschek reconnection differs qualitatively from the
non-relativistic one in case the reconnecting fields are not
exactly antiparallel. Due to very strong compression, a large
field may be built up in the outflow region so that a significant
fraction of the energy is ejected from the system as the Poynting
flux.

The article is organized as follows. The Sweet-Parker regime is
considered in sect.2. In sect. 3, the jump conditions at the
relativistic slow shocks are obtained. These conditions are
applied to the Petschek reconnection in sect. 4. Modification of
the picture in case the reconnection fields are not strictly
antiparallel are considered in sect.5; the detailed consideration
of this case is presented in Appendix. The results are summarized
in sect. 6.

\section{Relativistic Sweet-Parker reconnection}
In the Sweet-Parker configuration, the oppositely directed
magnetic fields are separated by a current sheet with a small
resistivity $\eta$. Some anomalous resistivity is assumed taking
into account that at high enough current density the sheet is
known to be unstable to the growth of tearing mode and other
instabilities. These instabilities are also developed in
relativistic current sheets (Zelenyi \& Krasnosel'skikh 1979;
Lyutikov 2003). In the spirit of the classical reconnection
models, we consider $\eta$ as a phenomenological parameter. The
resistive dissipation of the magnetic energy within the sheet may
be visualized as diffusion of the magnetic field towards the
neutral plane where the oppositely directed fields annihilate.
Outside the sheet, the magnetic field is frozen into the plasma
and therefore magnetic diffusion brings the plasma into the sheet.
The magnetic energy is converted into heat within the sheet and
the thermal pressure thrusts the plasma out of the sheet through
the sheet edges. The steady state is achieved when the plasma
inflow is balanced by the outflow from the edges of the sheet. The
reconnection rate may be roughly estimated from the integrated
balance of the energy and momentum within the sheet.

 Let the current sheet be in the $xz$ plane, the outer magnetic field be in the $x$
direction (Fig.1). The sheet width is $2\Delta$, the sheet length
is $2l$. The pressure equilibrium across the sheet implies
$$
p=\frac{B_0^2}{8\pi}, \eqno(1)
$$
where $p$ is the plasma pressure within the sheet, $B_0$ the
magnetic field outside the sheet. The hot plasma in the sheet
flows towards the edges of the sheet. The flow velocity in the
sheet, $v_{out}$, may be found from the momentum equation
$$
\frac{\partial}{\partial x}\left(wv_{\rm out}^2\gamma_{\rm
out}^2-p\right)=-j_zB_y,\eqno(2)
$$
where $w$ is the enthalpy of the plasma. Here and thereafter the
speed of light is taken to be unity. The current in the sheet is
estimated from Ampere's law as
$$
j_z=\frac{B_0}{4\pi\Delta}.\eqno(3)
$$
Substituting the $x$ derivative by $1/l$, one gets
$$
wv_{\rm out}^2\gamma_{\rm out}^2-p=\frac{B_yB_0}{4\pi\Delta}l,
 \eqno(4)
 $$
 The vertical component of the magnetic field,
 $B_y$, may be estimated from the flux conservation as $B_y\sim\Delta B_0/l$.
 Taking into account Eq.(1), one can see that the last term
 in the lhs of Eq.(4) is comparable with the rhs term.

\begin{figure}
\includegraphics[scale=0.6]{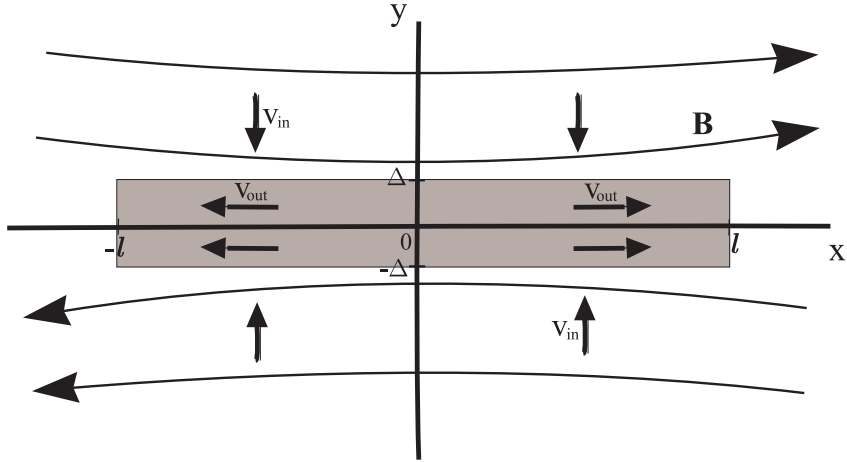}
 \caption{Geometry of the Sweet-Parker reconnection. The current sheet is shaded. Magnetic field lines are
 shown by thin arrows, the plasma velocities by thick arrows.}
\end{figure}

Let us consider the case when the magnetic energy density in the
inflow region exceeds the plasma rest energy density. Then the
 plasma in the sheet is relativistically hot and one can take
 $w=4p$. It follows immediately from Eq.(4) that the plasma may be accelerated
to high Lorentz factors only if $w$ significantly decreases
towards the edges of the sheet. According to Eq.(1), $p$ and,
consequently, $w$ are determined by the external magnetic field,
which does not decrease significantly along the sheet. Therefore
the plasma motion in the sheet is only mildly relativistic,
$\gamma_{\rm out}\sim 1$, $v_{\rm out}\sim 1$. Of course the
relativistically hot plasma may be eventually accelerated to high
Lorentz factors if the outer pressure considerably falls down.
However the reconnection rate is determined by plasma motion in
the region where the magnetic field is about its maximal value so
that the pressure in this part of the sheet does not change much.

Now let us consider the energy conservation. The energy influx is
$E_zB_0/4\pi=v_{\rm in}B_0^2/4\pi$, where $v_{\rm in}$ is the
inflow velocity determined just outside of the sheet where the
magnetic field is frozen into the plasma. The energy outflow is
$wv_{\rm out}$ so that the energy balance is written as
$$
v_{\rm in}\frac{B_0^2}{4\pi}l=4pv_{\rm out}\Delta.
 \eqno(5)
 $$
Making use of Eq.(1), one obtains
$$
\frac{v_{\rm in}}{v_{\rm out}}\sim\frac{\Delta}l,
 \eqno(6)
 $$
 so that the flow is roughly  incompressible (recall that all the above
 equations should be considered only as order of magnitude estimates).
Therefore even if the reconnecting field are not exactly
antiparallel and some small $z$-component of the magnetic field is
present in the inflow region, the picture remains the same because
$B_z$ remains small in the sheet and does not affect parameters of
the outflowing plasma.

In the steady state, ${\bf \nabla\times E}=0$, which implies
$\partial E_z/\partial y=0$ so that he electric field is the same
within and outside the sheet. Within the sheet, $E_z$ obeys Ohm's
law, which can be roughly written in the non-relativistic form,
$E_z=\eta j_z$, because the plasma motion within the sheet is only
mildly relativistic. Outside the sheet, $E_z=v_{\rm in}B_0$.
Eliminating $E_z$ and making use of Eq.(3), one can estimate the
inflow velocity as
 $$
v_{\rm in}=\frac{\eta}{4\pi\Delta}.
 \eqno(7)
  $$
Eliminating $\Delta$ from Eqs. (6) and (7), one finds finally
$$
v_{\rm in}= S^{-1/2};\qquad S\equiv 4\pi l/\eta\gg 1.
 \eqno(8)$$
So, contrary to what Blackmann \& Field (1994) and Lyutikov \&
Uzdensky (2003) claimed, the inflow velocity in the relativistic
Sweet-Parker regime remains much less than the speed of light.
This is because the flow velocity in the sheet remains mildly
relativistic so that the reconnection rate is not enhanced by the
Lorentz contraction.

\section{Jump conditions at the relativistic slow shock}

It was noticed by Petschek (1964) that the magnetic energy may be
liberated not only in current sheets but also at slow shocks. Let
us consider jump conditions at the slow shock in case the magnetic
energy exceeds the plasma energy. Let the upstream flow be cold,
$w_1=\rho_1$, whereas the downstream flow relativistically hot,
$w_2=4p_2$. In the frame of reference where the shock is at rest
and the upstream flow is perpendicular to the shock plane,
conservation of the energy and momentum fluxes are written as
$$
\rho_1\gamma_1^2v_{1}+\frac{B_{t1}E_t}{4\pi}=
w_2\gamma_2^2v_{n2}+\frac{B_{t2}E_t}{4\pi}; \eqno(9)
$$
$$
\rho_1\gamma_1^2v_{1}^2+\frac{B_{t1}^2}{8\pi}=
w_2\gamma_2^2v_{n2}^2+\frac{B_{t2}^2}{8\pi}+p_2; \eqno(10)
$$
$$
-\frac{B_nB_{t1}}{4\pi}=w_2\gamma_2^2v_{n2}v_{t2}-\frac{B_nB_{t2}}{4\pi}.
\eqno(11)
$$
Here the plasma density, $\rho$, and the enthalpy, $w$, are
measured in the plasma rest frame whereas the electro-magnetic
fields in the shock frame. Subscripts $n$ and $t$ refer to the
normal and tangential components, correspondingly, and we take
into account continuity of $B_n$ and $E_t$. The last is written as
$$
E_t= v_{1}B_{t1}=v_{n2}B_{t2}-B_nv_{t2}.\eqno(12)
$$

Shocks arising in the Petchek model are close to the switch-off
shocks for which $B_{t2}=0$. Recently Komissarov (2003)
demonstrated that, contrary to what was claimed before, such
shocks are evolutionary not only in the non-relativistic but also
in the relativistic case. For such shocks Eq.(12) yields
$$
v_{t2}=-\frac{B_{t1}}{B_n}v_{1}. \eqno(13)
$$
Substituting this relation into Eq.(11) and eliminating $w_2$ with
the aid of Eq.(9), one gets
$$
v_1^2=\frac{B^2_n(1-v^2_1)}{4\pi\rho_1+B^2_{t1}(1-v_1^2)},\eqno(14)
$$
which means that the upstream flow moves with the Alfven
velocity\footnote{In the plasma rest frame, the Alfven velocity is
$v_A=B'_n/\sqrt{4\pi\rho+B'^2}$, where $\bf B'$ is the magnetic
field in the plasma rest frame. One gets Eq.(14) taking into
account that in the frame moving with the Alfven velocity,
$B_n=B'_n$, $B_t=B'_t(1-v^2_A)^{-1/2}$.} as it should be.
Introducing the magnetization parameter
$$
\sigma\equiv\frac{B_1^2}{4\pi\rho_1\gamma_1^2},\eqno(15)
$$
one can see that
$$
v_1=\tan\theta\qquad{\rm at}\quad \sigma\gg 1,\eqno(16)
$$
where $\theta$ is the angle between the magnetic field and the
shock plane. This assumes $\theta<\pi/4$, which is fulfilled in
the Petchek picture.  In the opposite case the upstream velocity
may be found from the full biquadratic equation (14); if $B^2_1\gg
4\pi\rho_1$, $\theta>\pi/4$, the upstream flow is
ultra-relativistic such that $\sigma$ is not large. Below we
assume $\theta<\pi/4$.

Substituting Eq. (14) into Eq.(13) yields
$$
v_{t2}=-\cos\theta\sqrt{\frac{\sigma}{1+\sigma\cos^2\theta}}=-\left(1-\frac
1{2\sigma\cos^2\theta}\right).\eqno(17)
$$
The last equality is obtained in the limit $\sigma\gg 1$.
Substituting Eqs. (16, 17) into Eqs. (9, 10), one can easily find
in the same limit
$$
v_{n2}=\frac{\sin\theta}{2\sigma\cos^3\theta};\eqno(18)
$$
$$
\gamma_2=\sqrt{\sigma}\,\cos\theta;\eqno(19)
$$
$$
p_2=\frac{B^2_1\cos^2\theta}{8\pi}.\eqno(20)
$$
So the downstream flow is highly relativistic and directed at the
angle $\sim 1/\sigma$ to the shock plane.

Making use of the continuity equation,
$\rho_1\gamma_1v_{1}=\rho_2\gamma_2v_{n2}$, one finds the
downstream density
$$
\rho_2=2\rho_1\cos^2\theta\sqrt{\frac{\sigma}{\cos 2\theta}}.
\eqno(21)
$$
So the downstream flow is highly compressed. The downstream
temperature (or in fact the energy per particle) is easily found
from the equation of state:
$$
\frac Tm=\frac{\cos^2\theta}{4}\sqrt{\frac{\sigma} {\cos
2\theta}},\eqno(22)
$$
where $m$ is the average particle mass ($m=m_p/2$ in the
electron-proton plasma). Because $T\gg m$, the downstream flow may
be loaded by pairs; in this case $m$ becomes a function of the
temperature.

\section{Relativistic Petschek reconnection}

In the Petschek (1964) picture, the current sheet is localized at
$-l<x<l$, $y=0$ and pairs of slow shocks stem from the edges of
the sheet as far as the outer boundary of the box at $x=\pm L$
(Fig.2). At the shocks, the $x$ component of the magnetic field
drops to zero and the magnetic tension pushes the plasma away from
the reconnection region along the $x$ axis. In response to rapid
evacuation of the plasma from the reconnection region, the
upstream plasma is sucked in toward the $xz$ plane, together with
the upstream field.  The magnetic field in the inflow region is a
small perturbation to a uniform horizontal field,
$B_0\bf\widehat{x}$. The plasma pressure and inertia are
negligibly small at $\sigma\gg 1$; the electric force is also
small because the upstream velocity is non-relativistic, as it can
be checked {\it a posteriori}. Therefore the upstream field is
potential, ${\bf B}=\nabla\Phi$. According to the jump conditions,
the plasma upstream of the slow shocks moves with the Alfven
velocity, $v_{\rm in}=B_y/B_x\approx B_y/B_0$. It follows from the
above analysis that the angle between the shocks is only about
$1/\sigma$. Neglecting the inclination of the shocks, one should
solve Laplace's equation in the upper half-plane with the boundary
condition that the field becomes horizontal at large distances and
that the normal component of the field be $v_{\rm in}B_0$ between
$l$ and $L$ at the $x$-axis and, by symmetry, $-v_{\rm in}B_0$
between $-L$ and $-l$.

\begin{figure}
\includegraphics[scale=0.6]{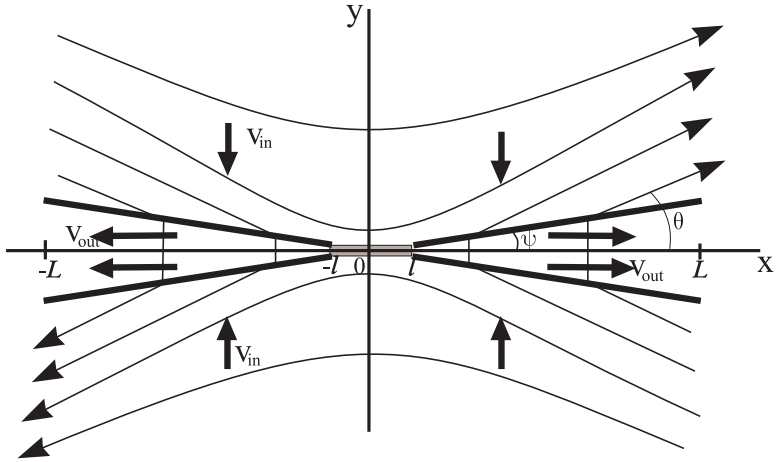}
\caption{Petschek reconnection. The slow shocks are shown by thick
lines; the rest elements are the same as in Fig.1}
\end{figure}

Following to Petschek (1964, see also Priest \& Forbes 2000), one
can write such a solution as
$$
\Phi=B_0x-\frac{v_{\rm
in}B_0}{\pi}\left\{\int_{-L}^{-l}\ln\left(y^2+(x-\xi)^2\right)d\xi\right.
$$
$$
\left.- \int_l^L\ln\left(y^2+(x-\xi)^2\right)d\xi\right\}.
\eqno(23)
$$
Then the field at the origin is
$$
B_x(0,0)=B_0\left(1-\frac{2v_{\rm in}}{\pi}\ln\frac
Ll\right).\eqno(24)
$$
Taking into account that the mechanism chokes itself off when the
field at the origin becomes too small, one can estimate a maximum
reconnection rate by putting $B_x(0,0)=0.5B_0$ to give
$$
v_{\rm in}=\frac{\pi}{4\ln L/l}.\eqno(25)
$$
The current sheet is described by the Sweet-Parker relation (8),
which may be written as $l=\eta/(4\pi v_{in}^2)$. Substituting
this into Eq.(25) and defining the Lundquist number via the
external scale, $S\equiv 4\pi Lc/\eta$, one gets finally
$$
v_{\rm in}=\frac{\pi}{4\ln S}.\eqno(26)
$$
So the maximal reconnection rate may be estimated as about 0.1 of
the speed of light. The reconnection rate does not approaches the
speed of light, contrary to what was expected by Blackman \& Field
(1994), because in the outflow, the Lorentz contraction is
compensated by a small angle between the slow shocks. According to
Eq.(16), the inclination angle of the magnetic field in the inflow
zone is $\theta=v_{in}\sim 0.1$.

\section{The relativistic Petschek reconnection in case $B_z\ne 0$}

Downstream of the slow shocks, the plasma  is highly compressed.
Therefore if the reconnecting fields are not strictly
antiparallel, the component of the magnetic field $B_z$ parallel
to the current
 may become so large that the structure of the plasma outflow may be
significantly affected.

Let us assume that some small $B_{z1}\equiv\alpha B_0$ is
presented in the inflow region (note that $B_z$ has the same sign
in the upper and lower inflow regions). Then the upper and lower
parts of the reconnected magnetic field line diverge in the
$z$-direction. In this case, the shrinkage of the field line after
the reconnection is accompanied by stretching in the $z$-direction
of the segment located within the outflow. Therefore the outflow
should be separated from the inflow by rotational discontinuities
where the magnetic field lines turn appropriately. It will be
shown in Appendix that a pair of slow shocks should appear between
the rotational discontinuities in order to satisfy the condition
$v_z(y=0)=0$. The structure of the flow is shown in Fig.3. In this
section we only roughly estimate parameters of the outflow zone;
the full solution is given in Appendix.

\begin{figure}
\includegraphics[scale=0.6]{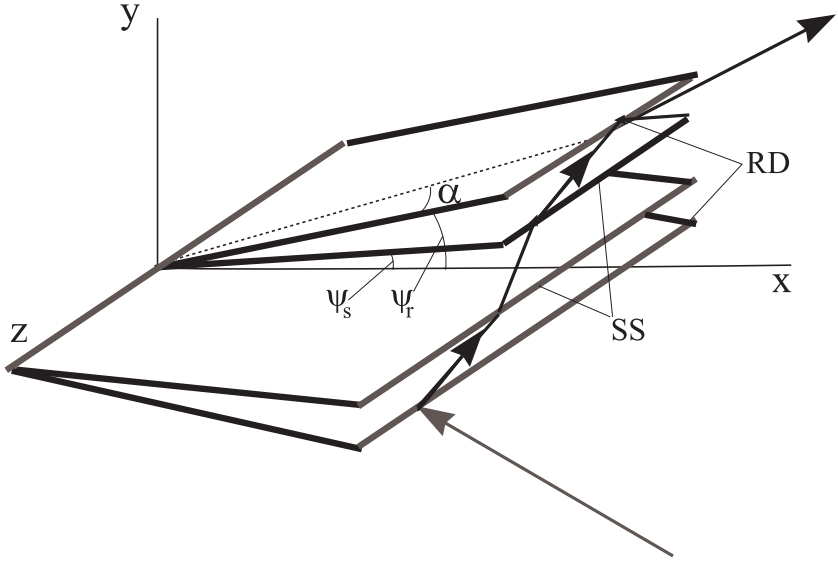}
\caption{Petschek reconnection in the oblique case. The current
flows along the $z$ axis; rotational discontinuities (RD) and slow
shocks (SS) stem from the current sheet. A typical magnetic field
line is shown by arrows. The trajectory of the point of
intersection between the field line and the rotational
discontinuity is shown by dotted line.}
\end{figure}

Let $\psi_r$ be the angle the rotational discontinuity makes with
the $xz$ plane. The magnetic field line shrinks with the speed of
light therefore the point of intersection between the field line
and the upper rotational discontinuity move in the $z$ direction
with the velocity $\alpha$ and in the $y$ direction with the
velocity $\psi_r$. At the bottom discontinuity, the corresponding
velocities have opposite signs therefore the segment of the field
line between the rotational discontinuities is inclined at the
angle $\alpha/\psi_r$ to the $xz$ plane. The $y$ component of the
magnetic field practically does not vary across the discontinuity.
Therefore the $z$-component of the field in the outflow may be
estimated as
$$
B_{z,\rm out}=
\frac{\alpha}{\psi_r}B_y=\frac{\alpha\theta}{\psi_r}B_0. \eqno(27)
$$
It is taken into account in the last equality that in the inflow
zone, the field line is inclined at a small angle $\theta$ to the
$xz$ plane (but $\theta\gg\psi_r$). The estimate (27) may be  from
the conservation of the magnetic flux crossing a fluid contour,
$\oint {\bf B\cdot v\times dl}=0$. Choosing a rectangular contour
in the $xy$ plane with the rotational discontinuity as a diagonal,
one finds $v_{\rm in}B_{z1}=B_{z,\rm out}\psi_r$. The inflow
velocity is equal to the Alfven velocity (16) because the
rotational discontinuity moves with respect to the plasma with the
Alfven velocity (see Appendix). Then one immediately gets Eq.(27).

The structure of the outflow is changed if the pressure of the
$B_z$ field in the outflow zone becomes sufficient to compensate
the outer pressure. In the rest frame of the outflow, the field is
equal to $B_{z,\rm out}/\gamma_{out}$, where $\gamma_{\rm out}$ is
the Lorentz factor of the plasma outflow. Taking into account that
the upstream field is not Lorentz transformed because it is
directed nearly parallel to the $x$ axis, one concludes that the
influence of $B_z$ is negligible if
$$
B_{z,\rm out}\ll B_0\gamma_{\rm out}.\eqno(28)
$$
In this case the outflow parameters are described by Eqs.(18) and
(28) like in the case $B_z=0$ so that one can write $\gamma_{\rm
out}=\sqrt{\sigma}$, $\psi=v_{n2}=\theta/(2\sigma)$. Taking this
into account and eliminating $B_{z,\rm out}$ from Eqs.(27, 28),
one can write the condition that $B_z$ in the outflow region is
dynamically insignificant as
$$
\alpha\ll (2\sqrt{\sigma})^{-1}.\eqno(29)
$$
In high-$\sigma$ plasmas, this condition is very restrictive; it
may be violated at rather small $\alpha$.

In the opposite limit, the pressure and the energy in the outflow
region are dominated by the magnetic field. The pressure
equilibrium in the transverse direction yields (cf. Eq.(28))
$$B_{z,\rm out}=B_0\gamma_{\rm out}.\eqno(30)$$
Increasing of the magnetic field at the rotational discontinuity
does not contradict to the general principles because the field
strength should remain the same only in the zero electric field
frame. In the laboratory frame, the field grows significantly. The
energy balance reads as
$$
v_{\rm in}B_0^2=B_{z,\rm out}^2\psi_r.\eqno(31)
$$
Taking into account Eq.(27), one gets the half-angle between the
rotational discontinuities as
$$
\psi_r=\theta\alpha^2.\eqno(32)
$$
and the Lorentz factor of the outflow as
$$
\gamma_{\rm out}=1/\alpha.\eqno(33)
$$
One can see that at the condition opposite to that of Eq.(29), the
opening angle of the outflow is larger whereas the outflow Lorentz
factor is less than the corresponding quantities obtained at
$B_z=0$. The transition to the $B_z=0$ case occurs when $\alpha$
is about the rhs of Eq.(29).

So even a small $B_z$ in the upstream flow significantly affects
the structure of the outflow. At the condition reverse to that of
Eq.(29) (but still $\alpha\ll 1$), the strongly magnetized
($B_{z,\rm out}\sim B_0/\alpha$) plane jet of the angular width
(32) is ejected with the Lorentz factor (33). In contrast with the
$B_z=0$ case, the energy flux is dominated by the Poynting flux.
However the reconnection velocity is equal to the Alfven velocity
in any case. Therefore the estimate (26) for the reconnection rate
remains valid even if $B_z\ne 0$.

\section{Conclusion}
The above analysis demonstrates that the classical reconnection
models may be straightforwardly generalized to the relativistic
regime. This regime arises when the energy density in the inflow
region is dominated by the magnetic field. Then the Alfven
velocity is close to the speed of light.  The reconnection rate in
this case may be estimated by substituting $v_A=c$ in the
classical formulas. The ejected plasma is relativistically hot so
that the pair production is in principle possible.

In the Sweet-Parker configuration, the plasma is pushed away with
mildly relativistic velocities. In the Petschek configuration, the
plasma is ejected with a large Lorentz factor within a very narrow
angle. The qualitative difference with the non-relativistic
reconnection arises if the reconnecting fields are not strictly
antiparallel. At the non-relativistic slow shock, the compression
ratio is finite so that if $B_z$ is small in the inflow region, it
remains small and does not affect the outflow parameters. The
plasma compression in the relativistic Petschek reconnection is
very high therefore even at a rather small $B_{1z}$ the
magnetization of the outflow may be very high so that most of the
energy of the reconnecting fields is taken away by the Poynting
flux. Therefore the relativistic Petschek reconnection should not
be considered as a mechanism for the direct conversion of the
magnetic energy into the plasma energy. However one can speculate
that the ejected energy is eventually transferred to the plasma in
the course of the flow expansion and/or development of MHD
instabilities.

\section*{Acknowledgment}
I am grateful to the anonymous referee for constructive criticism
and to David Eichler for valuable discussions and encouragement.
The work was supported by a grant from the Israeli-US Binational
Science Foundation and by a Center of Excellence grant from the
Israeli Science Foundation.

\section*{Appendix. Relativistic Petschek reconnection in the
oblique case}

Let the magnetic field in the inflow zone have a nonzero component
$B_z=\alpha B_0$. It was shown in Sect. 5, that one can neglect
the obliquity at the condition (29). In this Appendix, the
opposite case
$$
\alpha^2\sigma\gg 1\eqno(A1)
$$
will be considered. The structure of the flow is shown in Fig. 3.
The outflow is confined within two rotational discontinuities.
Inside the outflow, two slow shocks occur. The rotational
discontinuities and slow shocks are inclined to the $xz$ plane by
the angles $\psi_r$ and $\psi_s$, correspondingly. At the
rotational discontinuity, the magnetic field line turns whereas
the plasma is pushed in the $x$ direction. As we will see, the
jump conditions require nonzero $v_z$ beyond the discontinuity
whereas by symmetry, the plasma should move strictly parallel to
the $x$ axis at $y=0$. Therefore a pair of slow shocks arises
where the velocities are adjusted appropriately.

\begin{figure}
\includegraphics[scale=0.6]{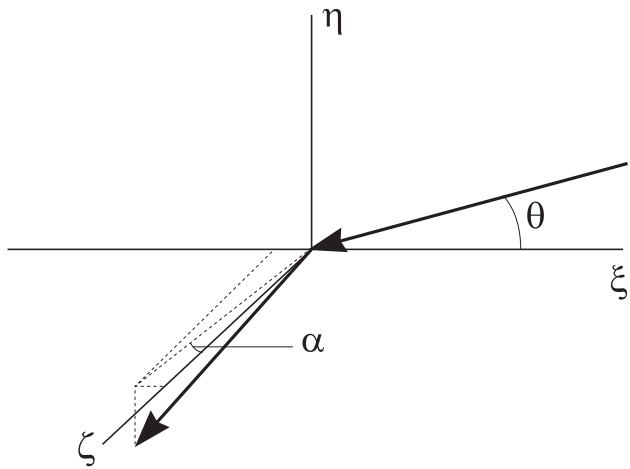}
\caption{Turn of the magnetic field at the rotational
discontinuity. The discontinuity lies in the $\xi\zeta$ plane. The
magnetic field is shown by arrows. }
\end{figure}

Let us first find the jump conditions at the rotational
discontinuities. Let us choose the coordinate system $(\xi,\eta,
\zeta)$ such that the upstream magnetic field lies in the
$\xi\eta$ plane and the discontinuity in $\xi\zeta$ plane (Fig.
4). This coordinate system is turned with respect to that in Fig.
3 by the angle $\alpha$ around the $y$ axis and by the angle
$\psi_r$ around the $z$ axis. Taking into account that
$\psi_r\ll\alpha,\theta$, one can neglect the tangential component
of the upstream velocity. Then the continuity of the tangential
component of the electric field reads as
$$
E_{\zeta}=v_1B_{\xi1}=v_{\eta2}B_{\xi2}-v_{\xi2}B_{\eta};\eqno(A2)
$$
$$
E_{\xi}=0=v_{\zeta2}B_{\eta}-v_{\eta2}B_{\zeta2}.\eqno(A3)
$$
Here we take into account continuity of $B_{\eta}$.

It is known from the general theory that the proper density and
entropy do not change at this sort of discontinuities. Employing
this fact from the beginning, we need only three equations in
order to find the rest of parameters. One can conveniently choose
the continuity of the flow,
$$
\gamma_1v_1=\gamma_2v_{\eta2},\eqno(A4)
$$
and the $\eta\eta$ and $\eta\zeta$ components of the momentum
flux:
$$
B^2_{\xi2}+B^2_{\zeta2}-E_{\eta2}^2=B^2_{\xi1};\eqno(A5)
$$
$$
\rho\gamma^2_2v_{\eta2}v_{\zeta2}-\frac
1{4\pi}\left(B_{\eta}B_{\zeta2}+E_{\eta2}E_{\zeta}\right)=0,\eqno(A6)
$$
where
$$
E_{\eta2}=B_{\zeta2}v_{\xi2}-B_{\xi2}v_{\zeta2}=\frac{B_{\zeta2}B_{\xi1}}{B_{\eta}}v_1.\eqno(A7)
$$
In the last equality, $v_{\zeta2}$ and $v_{\xi2}$ were eliminated
with the aid of Eqs.(A2) and (A3).

 Substituting Eqs.(A2) and (A7) into
Eq.(A6) yields
$$
\rho\gamma_2^2v_{\eta2}^2=\frac{B_{\xi1}^2}{4\pi}\left(\tan^2\theta-v_1^2\right),\eqno(A8)
$$
where $\theta$ is the angle between the discontinuity and the
upstream magnetic field. Then with account of Eq.(A4), one gets
$$
v_1^2=\tan^2\theta\left(1-\frac 1{\sigma}\right),\eqno(A9)
$$
where $\sigma\gg 1$ is defined by Eq.(15). So the upstream
velocity is the same as in the aligned case therefore obliquity
does not affect the reconnection rate.

Now let us return to Eq.(A8) and eliminate, with the aid of
Eqs.(A2) and  (A3), $v_{\zeta2}$ and $v_{\xi2}$ from $\gamma_2$.
Then substituting $v_1$ from Eq.(A9) yields the equation for
$v_{\eta2}$:
$$
\left(\frac{B^2_{\zeta2}+B^2_{\xi2}}{B^2_{\eta}}+1+\frac
1{\tan^2\theta(1-\tan^2\theta)}\right)v_{\eta2}^2
$$$$
+2\frac{B_{\xi2}}{B_{\eta}}v_{\eta2}-\frac 1{\sigma}=0.\eqno(A10)
$$
In order to avoid cumbersome expressions, let us take into account
that in the case of interest $\vert B_{\xi2}\vert\ll \vert
B_{\zeta2}\vert$ and, as follows from the preliminary estimates
(see sect. 5) and will be confirmed below by rigorous evaluation,
$\vert B_{\eta}/B_{\zeta2}\vert\sim \theta\alpha\ll\theta$. Then
the solution of Eq.(A10) at the condition (A1) is written as
$$
v_{\eta2}=-\frac{B_{\eta}}{B_{\zeta2}}\left(1+
\sqrt{\frac{B^2_{\xi2}}{B^2_{\zeta2}}+\frac
1{\sigma}}\right).\eqno(A11)
$$
The magnetic field downstream of the discontinuity is
perpendicular to the $x$ axis\footnote{Strictly speaking, the
magnetic field line should be perpendicular to the $x$ axis only
at $y=z=0$ (by symmetry) that is only between the slow shocks.
Above the shock, the magnetic field line lies in the plane set by
the normal to the shock and the field line between the shock.
Taking into account that $\psi_s$ is much less than all other
angles involved, one can safely assume that the field is
perpendicular to the $x$ axis already downstream of the rotational
discontinuities.} (see Fig. 3). Therefore
$$B_{\xi2}=-\alpha B_{\zeta2}.\eqno(A12)
$$
Then Eqs. (A11), (A2) and (A3) yield  at the condition (A1):
$$
v_{\eta2}=-\frac{B_{\eta}B^2_{\xi2}}{B^2_{\zeta2}};\qquad
v_{\zeta2}=2\alpha;\qquad v_{\xi2}=-1+2\alpha^2.\eqno(A13)
$$
Substituting Eqs.(A7), (A9) and (A12) into Eq.(A5),one finds
$$
B_{\xi1}=-\alpha B_{\zeta2},\eqno(A14)
$$
and then
$$
v_{\eta2}=-2\theta\alpha^2;\qquad \gamma_2=\frac
1{2\alpha^2}.\eqno(A15)
$$

Now let us return to Fig. 3 and the coordinate system $xyz$. It
follows form Eq.(A14) that in the outflow, $B_z=B_0/\alpha$.
Projecting $\bf v_2$ onto the $x$ axis, one finds
$v_{x2}=1-\alpha^2/2$ so that the outflow moves in the $x$
direction with the Lorentz factor $\gamma_{out}=1/\alpha$ in
accord with the estimate (33). By symmetry, the flow should move
exactly in the $x$ direction at $y=0$. However projections of $\bf
v_2$ on the $y$ and $z$ axis are non zero. Therefore a pair of
slow shocks arise in the outflow in order to adjust the velocities
appropriately.

Let us now find parameters of the slow shocks necessary to make
the plasma move exactly in the $x$ direction. Note that the
quantities upstream of the shock are the same as downstream of the
rotational discontinuity; we retain them with the index $2$. The
postshock quantities will be denoted by the index $3$. The jump
conditions may be conveniently found in the frame moving in the
$x$ direction with the Lorentz factor $\gamma_{out}$; the physical
quantities in this frame will be marked by prime (Fig. 5).  In
this frame, plasma moves only in the transverse direction with the
Lorentz factor
$$
\gamma'_2=\gamma_2/\gamma_{out}=1/2\alpha,\eqno(A16)
$$
whereas the shocks and rotational discontinuities move in the $y$
direction with the non-relativistic velocities
$$
V'_s=\psi_s\gamma_{out}=\psi_s/\alpha\quad {\rm and}\quad
V'_r=\psi_r/\alpha,\eqno(A17)
$$
correspondingly. The velocity $\bf v_2$ was calculated above with
respect to the rotational discontinuity. With respect to the
shock, the upstream normal velocity is
$$
v'_{n2}=v'_{\eta2}+V'_r-V'_s=-2\alpha\theta+(\psi_r-\psi_s)/\alpha.\eqno(A18)
$$
It follows from the estimates in sect. 5 and will be confirmed
below that the velocities of the shock and rotational
discontinuity are much less than the speed of light therefore the
full Lorentz transformation is not necessary.

\begin{figure}
\includegraphics[scale=0.4]{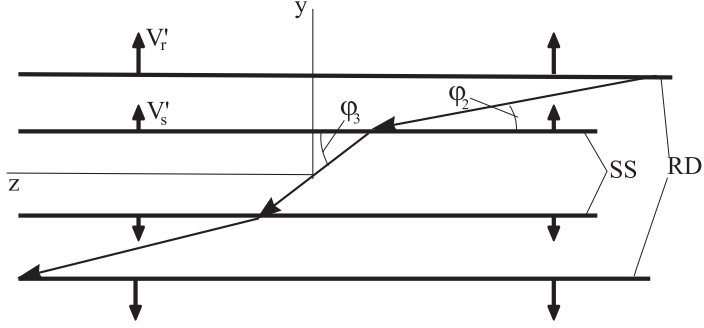}
\caption{Structure of the flow in the frame moving in the $x$
direction with the Lorentz factor $\gamma_{out}$, view end on. The
magnetic field line is shown by thin arrows, the rotational
discontinuities (RD) and slow shocks (SS) by thick lines,
velocities of the discontinuities by thick arrows. }
\end{figure}

Downstream of the shock, the plasma is at rest with respect to the
"primed" frame of reference therefore with respect to the shock,
tangential velocity is zero and the normal velocity is
$$
v'_3=-V'_s=-\psi_s/\alpha.\eqno(A19)
$$
The angle between the upstream magnetic field and the shock is
$$
\varphi_2=\left\vert\frac{B'_n}{B'_{t2}}\right\vert=
\left\vert\frac{B_n}{B_{t2}}\right\vert=-\frac{B_{\eta}}{B_{\zeta2}}=\theta\alpha.\eqno(A20)
$$
Continuity of the tangential component of the electric field and
of the normal component of the magnetic field take the form
$$
B'_{t2}(v'_{n2}-v'_{t2}\varphi_2)=B'_{t3}v'_{3};\eqno(A21)
$$
$$
B'_{t2}\varphi_2=B'_{t3}\varphi_3.\eqno(A22)
$$
Now the continuity of the energy and momentum fluxes may be
written as
$$
\rho_2\gamma^{'2}_2v'_{n2}=4p_3\gamma'_3v'_{3}+
\frac{B'_{t2}B'_{t3}}{4\pi}v'_3\left(\frac{\varphi_2}{\varphi_3}-1\right);\eqno(A23)
$$
$$
\rho_2\gamma^{'2}_2v^{'2}_{n2}=4p_3\gamma'_3v^{'2}_{3}+p_3+
\frac{B^{'2}_{t2}}{8\pi}v'_3\left(\frac{\varphi_2^2}{\varphi_3^2}-1\right);\eqno(A24)
$$
$$
\rho_2\gamma^{'2}_2v'_{n2}v'_{t2}+
\frac{B'_{n}B'_{t2}}{4\pi}\left(\frac{\varphi_2}{\varphi_3}-1\right)=0.\eqno(A25)
$$
Inspection of Eq.(A25), with account of Eqs.(A16, A18), shows that
$\varphi_3-\varphi_2\ll\varphi_2$ at the condition (A1). Moreover
one can safely neglect dynamical pressure as compared with the
thermal and magnetic pressure in Eq.(A24). Then one gets
$$
p_3=\frac{B^{'2}_{t2}}{4\pi}\left(1-\frac{\varphi_2}{\varphi_3}\right).\eqno(A26)
$$
Substituting this relation into Eq.(A23) and eliminating
$\rho_2\gamma^{'2}_2v'_{n2}$ with the aid of Eq.(A25) (one can put
$v_{t2}=1$ and $\gamma_3=1$ there), one finds
$$
\varphi_2=-3v'_{3},\eqno(A27)
$$
which yields, with account of Eqs.(A19, A20),
$$
\psi_s=\frac 13 \theta\alpha^2.\eqno(A28)
$$
Substituting Eq.(A27) into Eq.(A21), with account of Eq.(A20),
yields
$$
v'_{n2}=-(4/3)\varphi_2=-(4/3)\theta\alpha.\eqno(29)
$$
Now one finds from Eqs.(A18), (A28) and (A29) that
$$
\psi_r=\theta\alpha^2\eqno(A30)
$$
in accord with the estimate (32).

One can find from the Lorentz transformation that
$B'_n=B_n/\gamma_{out}=\theta\alpha B_0$; $B'_{t2}=\alpha
B_{z2}=B_0$. It follows from Eq.(A25) that
$$
1-\frac{\varphi_2}{\varphi_3}=\frac 1{3\alpha^2\sigma}.\eqno(A31)
$$
Then Eq.(26) yields
$$
p_3=\frac{B_0^2}{12\pi\alpha^2\sigma}.\eqno(32)
$$
Now one can find the ratio of the energy transferred away by the
plasma and by the electro-magnetic field:
$$
\frac{4p\gamma_{out}^2\psi_s}{B_z^2\psi_r/4\pi}=\frac
1{3\alpha^2\sigma}.\eqno(A33)
$$
So at the condition (A1), the energy of the reconnecting field is
transferred away predominantly in the form of the Poynting flux.

\end{document}